\documentclass[Crown,sagev,times]{sagej}

\usepackage{moreverb,url}

\usepackage[colorlinks,bookmarksopen,bookmarksnumbered,citecolor=red,urlcolor=red]{hyperref}

\newcommand\BibTeX{{\rmfamily B\kern-.05em \textsc{i\kern-.025em b}\kern-.08em
T\kern-.1667em\lower.7ex\hbox{E}\kern-.125emX}}

\def\volumeyear{2021}

\usepackage{subfigure}
\usepackage{pgfgantt}

\begin{document}

\runninghead{Elbakly et al.}

\title{Entropy-patch choked-nozzle interaction: quasi-steady and inertial modeling regimes mapped and limits of linearization established}

\author{Karim Elbakly\affilnum{1}, Steven J. Hulshoff\affilnum{2}, Friedrich Bake\affilnum{3}, Cornelis H. Venner\affilnum{1}, and Lionel Hirschberg\affilnum{1}}

\affiliation{\affilnum{1}University of Twente, Engineering Fluid Dynamics, PO Box 217, 7500 AE Enschede, The Netherlands\\
\affilnum{2}Delft University of Technology, Faculty of Aerospace Engineering, Kluyverweg 2,  2629HS Delft, The Netherlands\\\affilnum{3}Division for Acoustic and Electromagnetic Methods, Department of Non-destructive Testing, Bundesanstalt für Materialforschung und -prüfung
(BAM), Unter den Eichen 87, 12205 Berlin, Germany}

\corrauth{Karim Elbakly, University of Twente, Engineering Fluid Dynamics, PO Box 217, 7500 AE Enschede, The Netherlands}

\email{k.m.a.a.elbakly@student.utwente.nl}

\begin{abstract}
The effects of entropy-patch shape, size, and strength on the upstream acoustic response generated by entropy-patch choked-nozzle interactions are investigated. Numerical-simulation-based investigations, using a two-dimensional planar Euler code, reveal the existence of two distinct modeling regimes: the quasi-steady (matching-condition) regime and the inertial regime, respectively. The ratio of the entropy-patch streamwise length scale to the nozzle throat height was found to be an order parameter, which allows one to determine which of the two modeling regimes applies. Indeed, for entropy patches with a streamwise length scale smaller or equal to the nozzle throat height, the inertial model provides a satisfactory prediction of the upstream acoustic response. For entropy patches with a streamwise length scale larger than the nozzle throat height, the matching condition model has superior predictive accuracy. The entropy patch's shape was judged to have only a slight impact on the applicable modeling regime. Additionally, the study examined entropy-patch strength using the ratio of area-specific perturbation energy to area-specific upstream energy as an order parameter, establishing that both above-mentioned linear models are only valid for weak entropy patches. These findings provide a framework for selecting appropriate models for entropy-patch choked-nozzle interaction scenarios, furthering the fundamental understanding of indirect noise-driven combustion instability. 
\end{abstract}


\maketitle

\section{Introduction}\label{section:intro}
Engineering systems employing turbulent combustion usually have high levels of noise production, due to both direct \& indirect combustion-noise sources. Direct sources, due to 
unsteady gas expansion and contraction in the reactive region, have been widely studied \cite{strahle_1971,Dowling_2,MorgansReview,genot2021aluminum}. Indirect sources include entropy noise and vorticity noise. In particular, both entropy patches and vortices produce sound waves as they exit the combustion chamber through a nozzle or turbine. Some of these sound waves are radiated into the environment, whilst some are reflected back into the combustion chamber. The latter can produce new entropy patches and vortices, which in turn lead to the production of new sound waves as they exit the combustion chamber. Under unfavorable circumstances, this results in a feedback loop that promotes combustion instability or self-sustained pressure pulsations. Thermo-acoustic combustion chamber instabilities driven by indirect combustion noise are a potential issue in aero-engines and electrical-power generation turbines \cite{Dowling_2,MorgansReview}. In large solid rocket motors, vorticity-noise-driven self-sustained pressure pulsations are an established issue \cite{Dotson_5,Hulshoff_6,AnthoineJPP2002,HirschbergJASAEL,HirschbergAIAA1,HirschbergAIAA2}.

In order to cultivate a fundamental understanding of complex phenomena such as indirect combustion noise, it is standard practice to perform order-reduction by designing experiments in which only one effect is dominant---or on occasion, when the former has been done, a few effects.  \cite{AnthoineJPP2002,Bake_2009,KingsSpray_12,KingsThesis,HirschbergAIAA4,Domenico_2021,Noiray_2021,HirschbergEXIF,HirschbergJSV_2}. A prime example of this approach are Anthoine's et al.~\cite{AnthoineJPP2002} cold-gas (without combustion) scale-model experiments, which were used to investigate self-sustained pressure pulsations in solid-rocket motors. Indeed, these demonstrated the importance of the integrated nozzle's nozzle-cavity volume on indirect noise produced by essentially nonlinear azimuthal-vortex-nozzle (or ring-vortex-nozzle) interaction. Other examples are Bake's et al.~\cite{Bake_2009} canonical entropy-noise experiment, De Domenico's et al.~experiment \cite{Domenico_2021}, Noiray \& Wellemann's experiment~\cite{Noiray_2021}, and Hirschberg's et al.~entropy \& axial-component-vorticity noise experiments~\cite{HirschbergAIAA4,HirschbergEXIF,HirschbergJSV_2}. Moreover, the practice of studying indirect combustion-noise sources in isolation has also been successfully used for the development of analytical \& numerical indirect combustion-noise models \cite{MC,Ffowcs_Williams_howe_1975,LeykoEtAl,HirschbergAIAA2,GentilEtAl_AIAA}. 

Of the two indirect combustion-noise sources, entropy noise has been the most widely studied, as evidenced by the high number of citations of two seminal articles by Marble \& Candel \cite{MC} and Ffowcs Williams \& Howe \cite{Ffowcs_Williams_howe_1975}. Marble \& Candel's {one-dimensional} (1-D) modeling approach~\cite{MC,Ffowcs_Williams_howe_1975}, based on the notion of plane entropy-wave interaction with a nozzle, appears to be the most widely applied. In contrast, Ffowcs Williams \& Howe's modeling approach considers, three-dimensional patches of the fluid---with relative-excess mass---convected by the flow~\cite{Ffowcs_Williams_howe_1975}. 

Ffowcs Williams \& Howe seem to have argued that ``{\it to elicit in detail the physical mechanisms responsible for the generation of sound}'' \cite{Ffowcs_Williams_howe_1975} the inclusion of acceleration/unsteadiness is an ineluctable ingredient for a model. Whereas, Marble \& Candel astutely pointed out: ``{\it When the scale of the disturbance impinging upon the nozzle is large in comparison with the nozzle length ... the response of the nozzle is well approximated by a matching-condition analysis. Though limited in the range of frequency over which it is applicable, the results which follow from this approximation are simple and extremely useful. The idea is simply that, to disturbances of very long wavelength, the nozzle appears as a discontinuity in the state of the medium supporting the propagation; the state gradients ... become discontinuities. The nozzle then provides matching conditions, between uniform upstream and downstream states, which may be derived from conservation laws and the geometric description of the nozzle.}''
      
For the case of choked-nozzle-flow experiments, Hirschberg et al.~\cite{HirschbergJSV_2} used Marble \& Candel's above-quoted observation to formulate a bare-bones matching-condition model. Said model was validated by comparison with Leyko's et al.~\cite{LeykoEtAl} simulation results \cite{HirschbergJSV_2}. Moreover, Hirschberg et al.~\cite{HirschbergJSV_2} pointed out that in the cases where matching-condition modeling is applicable: sound production is due to a temporary axial mass-flow rate change caused by the passage of an entropy patch through the nozzle throat.      

Given that Ffowcs Williams \& Howe's \cite{Ffowcs_Williams_howe_1975} method is not limited to one-dimensionality, it allows for the investigation of the entropy-patch size on sound production. Ffowcs Williams \& Howe \cite{Ffowcs_Williams_howe_1975} investigated the influence of entropy-patch size on sound generation, which they termed ``{\it acoustic bremsstrahlung}'' or ``{\it bremsstrahlung}'' \cite{Ffowcs_Williams_howe_1975}. In particular, they used their model to compare the sound generation of a duct-sized entropy ``{\it slug}'' to that of a much smaller spherical ``{\it pellet},'' as these pass through a duct contraction or a nozzle \cite{Ffowcs_Williams_howe_1975}. One should note that Ffowcs Williams \& Howe only considered low-Mach-number flow; viz., they did not consider choked-nozzle flows.               

In the present study, inspired by Marbel \& Candel~\cite{MC} and Ffowcs Williams \& Howe's \cite{Ffowcs_Williams_howe_1975} work, but with a focus on choked-nozzle flows, we investigated the influence of an entropy-patch shape, size, and strength on the upstream-traveling acoustic response due to entropy-patch choked-nozzle interaction using numerical simulations. Linear reduced-order model-based scaling rule analysis was performed on the results. The fundamental questions which we aimed to answer were: 

\begin{enumerate}
    \item Under which conditions is the acceleration of an entropy patch an essential modeling ingredient? In other words, is there an inertial modeling regime?
    \item When does simple quasi-steady (matching-condition) modeling, which does not include the explicit modeling of acceleration, suffice? 
    \item Is there a dimensionless order parameter that allows one to determine if one finds oneself in either the inertial modeling or the matching-condition regime? 
    \item Is there a dimensionless order parameter that allows one to determine if linearization is an appropriate modeling strategy and when nonlinearity becomes essential? 
\end{enumerate}

Our work builds on Kowalski et al.~\cite{Kowalski,KowalskiMScThesis}, were a two-dimentonal (2-D) planar Euler code tailored for internal flow acoustics (EIA) developed by Hulshoff\cite{EIA}, was used to study pressure pulsations due to entropy-patch choked-nozzle interaction. Indeed, Kowalski et al.~\cite{Kowalski,KowalskiMScThesis} took the important first steps to answering the above-raised questions. However, their work can only be qualified as preliminary, as neither the observed order of accuracy nor the numerical error were quantified for the reported results. What's more the quantity of interest, the upstream acoustic response due to entropy-patch-nozzle interaction, was polluted by spurious acoustic modes. Indeed, the signal to noise ratio (SNR) was such that simulations with very small entropy patches could not producce clear results. Moreover, Kowalski et al.~\cite{Kowalski,KowalskiMScThesis} relied on rough estimations of key parameters; viz.,  excess mass, excess-mass density \& the steady convective acceleration of an entropy patch. 

In our study, we employed Elbakly's \cite{Elbakly_report} new and vastly improved numerical simulation strategy, which improved the SNR in simulation results. The improvement was such that we were able to simulate the interaction of patches much smaller than what had previously been possible. Moreover, we employed a new version of EIA, with functionalities developed by Elbakly~\cite{Elbakly_report}, which allows one to directly extract the excess mass, excess-mass density \& the steady convective acceleration of an entropy patch. The latter significantly improves the reduced-order model-based analysis. 

Additionally, Kowalski et al.~\cite{Kowalski,KowalskiMScThesis} did not find the appropriate dimensionless order parameters, which allow one to discriminate between the inertial \& matching-condition regimes nor between linear \& nonlinear regimes. In this text, we report suitable dimensionless order-parameters for the first time. 

Ulf Michel, to whom this article and this issue is dedicated: has---besides his many  scientific achievements in Aero-Acoustics---made a seminal contribution to the experimental investigation of indirect combustion noise---entropy noise in particular. Indeed, he was the instigator of the work on entropy patches and entropy noise at the DLR in Berlin-Charlottenburg, which led to the development of the now canonical Entropy Wave Generator (EWG) experiment \cite{FriedrichBakeUlfMichel,Bake_2009}. 

However, not only did Ulf Michel set himself apart through his indubitable scientific achievements, as an exceptionally sportive mentor to many young scientists throughout the years---including Friedrich Bake---he has left an indelible mark on the state-of-the-art.

\section{General Approach}

\begin{figure}[h]
    \centering
    \includegraphics[width=\linewidth]{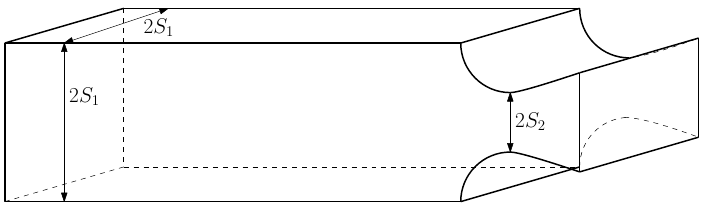}
    \caption{Computational domain representation of nozzle with upstream channel \cite{hirschberg_PHD}}
    \label{fig:computational_domain}
\end{figure}

To investigate the limits of the matching-condition \& inertial model, numerical simulations were carried out using EIA\cite{EIA}. The pressure pulsations, due to entropy-patch choked-nozzle interaction, obtained from the numerical simulations were scaled using the reduced-order models; viz., the matching-condition \& inertial models. The computational domain used throughout the study can be seen in Fig. \ref{fig:computational_domain}. The computational domain consists of a nozzle with a throat height of $2S_2$ and a depth of $2S_1$ connected to an upstream chamber with a channel height of $2S_1$ and depth $2S_1$. 

The reduced-order models are presented in \textbf{\S Reduced-order models}. The computational procedure which was used is presented in \textbf{\S Computational procedure}.

\section{Reduced-order models}\label{sec:reduce_order_models}

In this section, two bare-bones models are derived for the prediction of an upstream acoustic pressure response due to an entropy patch interacting with a choked-nozzle. The first model will be referred to as the matching-condition model, it is based on the quasi-steady analysis of Marble \& Candel \cite{MC}. The second model, inspired by Ffowcs Williams \& Howe's analytical model \cite{Ffowcs_Williams_howe_1975}, will be referred to as: the inertial model.

\subsection{Matching condition model}

Here the derivation of a matching condition model, for the prediction of upstream acoustic pressure response $p^\prime_u$, due to the interaction of an entropy patch of excess density $\rho^\prime_e=\rho_e-\rho_u$ with a chocked nozzle is provided, were $\rho_u$ is the density of the upstream channel and $\rho_e$ is the density of the entropy patch. Upstream from the nozzle, the flow is taken to be one-dimensional (1-D) viz.; the local flow variables vary only in the streamwise direction. Additionally, it is assumed that $\rho^\prime_e/\rho_u$ is small enough, such that the entropy patch is carried by the base flow without affecting it. Lastly, it is assumed that the interaction time scale of the entropy patch with the nozzle is significantly larger than the travel time of a material element through the nozzle. 

Now, let us consider d’Alembert’s solution to the one-dimensional wave equation: 
\begin{align}
    p^{\prime}=p^{+}(x-(c+u)t)+p^{-}(x+(c-u)t)\\
    u^{\prime}=u^{+}(x-(c+u)t)+u^{-}(x+(c-u)t)
\end{align}

\noindent where $p^\pm$ and $u^\pm$ are pressure \& velocity perturbations (the subscripts $+$ and $-$ denote downstream and upstream traveling waves) at a position $x$ at time $t$ and $c$ is the sound speed. If one considers an infinite duct with anechoic terminations and only perturbations traveling in the upstream direction, the equations reduce to the following:

\begin{align}
    p^{\prime}=p^{-}(x+(c-u)t)\\
    u^{\prime}=u^{-}(x+(c-u)t)
\end{align}

\noindent Appling the above equations to the linearized one-dimensional momentum conservation equation

\begin{align}
    \rho_u\frac{\partial u^{\prime}}{\partial t}=-\frac{\partial p^{\prime}}{\partial x}
\end{align}

\noindent one finds that the upstream pressure perturbation $p^\prime_u$ can be written as follows:

\begin{align}\label{eq:relation_purt_pressure_velocity}
    p^\prime_u = - \rho_uc_uu^\prime_u
\end{align}

The upstream velocity perturbation $u^\prime_u$ can be found by considering the upstream Mach number:

\begin{align}
    M_u = \frac{u_u}{c_u}
\end{align}

\noindent which is constant for a choked-nozzle flow. Thus, taking the total derivative

\begin{align}
    \mathrm{d}(M_u) = \mathrm{d}\left(\frac{u_u}{c_u}\right) = 0
\end{align}

\noindent one finds the following relation

\begin{align}\label{eq;relation_velocity_speedsound}
    \frac{u^\prime_u}{u_u}=\frac{c^\prime_u}{c_u}.
\end{align}

\noindent Assuming the fluid in the system to be a perfect gas, the speed of sound can be written as 

\begin{align}
    c_u^2=\gamma\frac{p_u}{\rho_u}
\end{align}

\noindent where $\gamma=c_p/c_v$ is the ratio of heat capacities at constant pressure and volume, respectively. Taking the total derivative of this expression yields 

\begin{align}
    \mathrm{d}( c_u^2) = \frac{\gamma}{\rho_u}\;\mathrm{d}(p_u)+\gamma p_u\;\mathrm{d}\left(\frac{1}{\rho_u}\right)
\end{align}

\noindent If one assumes isobaric generation of an entropy patch, one finds:  

\begin{align}
    2 \frac{c^\prime_u}{c_u} = - \frac{\rho_u^\prime}{\rho_u} = -\frac{\rho^\prime_e}{\rho_u}
\end{align}

\noindent Using this relation and Eq. \ref{eq;relation_velocity_speedsound} to rewrite the velocity perturbation in Eq. \ref{eq:relation_purt_pressure_velocity}, allows one the find:

\begin{align}\label{eq:matching_conditon_model}
    \boxed{p^\prime_u = \frac{1}{2}\rho_uc_uu_u\left(\frac{\rho^\prime_e}{\rho_u}\right) = \frac{1}{2}\rho_uc^2_uM_u\left(\frac{\rho^\prime_e}{\rho_u}\right) = \frac{1}{2}\gamma p_uM_u\left(\frac{\rho^\prime_{e}}{\rho_u}\right)}
\end{align}

\noindent where $p_u = \rho_u c^{2}_u/\gamma$ is the static pressure of the upstream channel. Hereon, this quasi-steady model will be referred to as the matching-condition model.

\subsection{Inertial model}
In this section, the derivation of an inertial model inspired by Ffowcs Williams \& Howe's analytical model~\cite{Ffowcs_Williams_howe_1975} is provided.

\begin{figure}[h]
    \centering
    \includegraphics[width=0.85\linewidth]{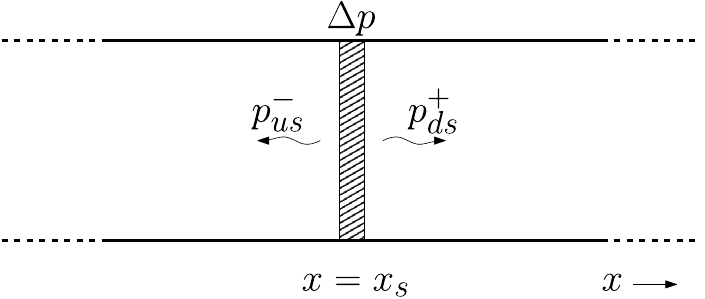}
    \caption{Unsteady pressure discontinuity $\Delta p$ at $x=x_s$, in a uniform 1-D ducted flow emanating plane acoustic waves $p_{us}^-$ and $p_{ds}^+$. }
    \label{fig:pressure_disc_straight_channel}
\end{figure}

Consider a sound source in the form of an unsteady pressure discontinuity $\Delta p$ at a position $x = x_\mathrm{s}$ (Fig. \ref{fig:pressure_disc_straight_channel}). It is assumed that said pressure discontinuity is present in a duct with a uniform cross-section and 1-D flow with Mach number $M$. Given that the pressure discontinuity is unsteady: pressure waves emanate from it in the upstream $p_{us}^-$ and downstream $p_{ds}^+$ direction.   Here the subscripts $us$ and $ds$ indicate perturbations upstream and downstream relative to the sound source. The superscripts $-$ and $+$ denote downstream and upstream traveling waves, respectively. 

Assuming an infinite duct with anechoic terminations, $\Delta p $ can be expressed in terms of the emitted pressure perturbations as follows:

\begin{equation}\label{eq:delta_p_def}
\Delta p = p_{ds}^+ - p_{us}^-.
\end{equation}

\noindent The mass flux across the pressure discontinuity is conserved, and thus one can write

\begin{equation}\label{eq:mass_flux_conservation}
(\rho u)_{ds}^{\prime}=(\rho u)_{us}^{\prime}
\end{equation} 

\noindent which can be re-written as

\begin{equation}\label{eq:mass_flux_rd}
\rho_{ds}^+u+\rho u_{ds}^+ = \rho_{us}^-u + \rho u_{us}^-
\end{equation} 

\noindent where the terms without sub or superscripts are the mean flow variables (higher-order perturbation terms are neglected). 

For an isentropic flow, the density perturbations can be expressed as:

\begin{equation}\label{eq:isentropic_relation_density_pressure}
\rho^\pm = \frac{p^\pm}{c^2}
\end{equation} 

\noindent Using d’Alembert’s solution to the one-dimensional wave equation and the linearized one-dimensional momentum conservation equation, the velocity perturbations can be expressed as 

\begin{equation}\label{eq:velocity_pertubation}
u^\pm = \pm\frac{p^\pm}{\rho c}
\end{equation}

\noindent Using Eq. \ref{eq:isentropic_relation_density_pressure} \& Eq. \ref{eq:velocity_pertubation} to re-write the perturbation terms in Eq. \ref{eq:mass_flux_rd}, one finds 

\begin{equation}\label{eq:relation_pressure_pertubations}
    p_d^+(1+M_s)=p_u^-(-1+M_s)
\end{equation}

\noindent where $M_\mathrm{s}$ is the Mach number at the sound source. Using Eq. \ref{eq:delta_p_def} the pressure perturbations can be written as 

\begin{align}
    p_{us}^{-}&=-\frac{1+M_{s}}{2}\Delta p\label{eq:p_u_minus}\\
    p_{ds}^{+}&=\frac{1-M_{s}}{2}\Delta \label{eq:p_d_plus} p
\end{align}

\noindent This result will be applied to the subsonic parts of a choked-nozzle with varying cross-sectional area $A = A(x)$. Here it is assumed that the nozzle is quasi-1-D; viz., the rate of change of the cross-sectional area is considered to be very small\cite{thompson1972compressible}. 

\begin{figure}
    \centering
    \includegraphics[width=0.85\linewidth]{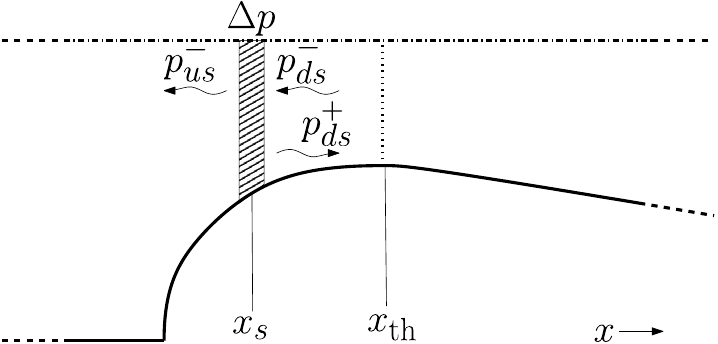}
    \caption{Acoustic pressure waves emanating from a fluctuating pressure discontinuity $\Delta p$ located at $x=x_s$ in the converging part of a choked-nozzle. As the nozzle is choked and the flow is 1D, one has sonic line at $x=x_\mathrm{th}$, viz., at the throat. }
    \label{fig:pressure_disc_nozzle}
\end{figure}

At the nozzle throat, we assume that downstream traveling waves are reflected.  Thus, an additional upstream traveling perturbation $p^-_{ds}$ is to be accounted for downstream from $\Delta p$ (Fig. \ref{fig:pressure_disc_nozzle}).  The reflection coefficient at $x=x_{th}$ can be take as\cite{MC}

\begin{equation}
    R=\frac{1-\frac{\gamma-1}{2}M_u}{1+\frac{\gamma-1}{2}M_u}
\end{equation}

\noindent Thus, $p^\prime_{us}$ can be expressed as follows:

\begin{equation}
    p_{us}^{\prime}=p_{us}^-+p_{ds}^-=p_{us}^-+Rp_{ds}^+ 
\end{equation}

\noindent which using Eq. \ref{eq:p_u_minus} and Eq. \ref{eq:p_d_plus} can be expressed as follows:

\begin{equation}
    p_{us}^{\prime}=-\frac{\Delta p}{2}\left((1+M_\mathrm{s})-R\left(1-M_\mathrm{s}\right)\right).
\end{equation}

To find the pressure perturbation observed in the upstream channel due to an upstream traveling pressure wave $p^\prime_u$ at location $x = x_u$, one must account for the change in cross-sectional area $A$. The acoustic power emitted directly upstream of the sound source can be taken as (\textbf{\S Appendix})

\begin{equation}\label{eq:acoustic_power_source}
|\Phi_s^-| = \frac{A_s}{\rho_sc_s} |p_{us}'|^2\left(1 - M_s\right)^2
\end{equation} 

\noindent and at the observer position, the acoustic power can be expressed as

\begin{equation}\label{eq:acoustic_power_obs}
|\Phi_u^-| = \frac{A_u}{\rho_uc_u} |p_u'|^2\left(1 - M_u\right)^2.
\end{equation}

Given that we consider an isentropic system, it must hold true that $|\Phi_s^-|=|\Phi_u^-|$. This allows one to express the magnitude of the observed pressure perturbation as:

\begin{equation}\label{eq:p_ob_general}
|p_u'| = \sqrt{\frac{\rho_u c_u}{\rho_s c_s}\frac{A_s}{A_u}}\left(\left(1 + M_s\right)-R\left(1 - M_s\right)\right)\left(\frac{1-M_s}{1-M_u}\right)\frac{\Delta p}{2}.
\end{equation} 

Noting that  an unsteady force from a wall on the fluid is a sound source \cite{Curle,Pierce}, we write

\begin{equation}\label{eq:delta_p_f_a}
\Delta p = \frac{F_\mathrm{x}}{A_\mathrm{s}}
\end{equation}

\noindent where the force $F_\mathrm{x}$ is exerted by the walls of the nozzle inlet on the fluid. $F_\mathrm{x}$ is due to the acceleration of an entropy patch, which is taken to be a point particle with an excess mass $m_e$. i.e., one can write

\begin{equation}\label{eq:fx}
F_x = m_e\left(u\frac{\mathrm d u}{\mathrm d x}\right).
\end{equation}

Using the above, Eq. \ref{eq:p_ob_general} can be re-written as follows:

\begin{equation}\label{eq:p_ob_specific}
\boxed{
|p_u'| = \sqrt{\frac{\rho_u c_u}{\rho_s c_s}\frac{1}{A_uA_s}}\left(\left(1 + M_s\right)-R\left(1 - M_s\right)\right)\left(\frac{1-M_s}{1-M_u}\right)\frac{m_e}{2}\left(u\frac{\mathrm d u}{\mathrm d x}\right)}
\end{equation} 

Using Bernoulli’s principle and isentropic perfect gas relations\cite{thompson1972compressible}, the $\sqrt{\rho_u c_u/\rho_s c_s}$ and $A_s$ terms in the above equation can be expressed as follows: 

\begin{equation}\label{eq:rho_c_u_over_rho_c_s}
\sqrt{\frac{\rho_u c_u}{\rho_s c_s}} = \left(\frac{1+\frac{\gamma-1}{2}M_s^2}{1+\frac{\gamma-1}{2}M_u^2}\right)^{\frac{\gamma+1}{4(\gamma-1)}}
\end{equation}

\noindent and 

\begin{equation}\label{eq:A_s}
A_s= \frac{A_\mathrm{th}}{M_s}\left(1+\frac{\gamma-1}{\gamma+1}(M_s^2-1)\right)^{\frac{\gamma + 1}{2(\gamma-1)}}.
\end{equation}

where the cross-sectional area of the throat is taken to be $A_\mathrm{th} = 4S_1S_2$ (Fig. \ref{fig:computational_domain}). Going forward, Eq. \ref{eq:p_ob_specific} will be referred to as the inertial model.

\section{Computational procedure}\label{sec:computationl_procedure}
Parametric studies of entropy-patch-choked-nozzle interaction were carried out using Hulshoff's two-dimensional Euler Internal Aeroacoustics code (EIA) \cite{EIA}. EIA makes use of the compressible lossless
governing (Euler) equations given by 
\begin{align}
 \frac{\partial \rho}{\partial t}+\nabla\cdot\left(\rho\mathbf u\right)&= 0\label{eq:MassEuler}\\
 \frac{\partial \rho\mathbf u}{\partial t}+\nabla\cdot\left(\rho\mathbf u\mathbf u+p\mathbf I\right)&= \rho\mathbf F_E\label{eq:MomentumEuler}\\
 \frac{\partial  E_T}{\partial t}+\nabla\cdot\left(( E_T+p)\mathbf u\right) &=  Q_E \label{eq:EnergyEuler}
\end{align}
to resolve the domain, where $ E_T=\rho (e+|\mathbf u|^2/2)$ is the total energy density, $\rho\mathbf F_E$ is an external momentum source density, and $Q_E$ is an external energy source.  $\mathbf F_E$ can be used to generate vortices as previously done by Hulshoff et al. \cite{Hulshoff_6} \& Hirshberg et al. \cite{HirschbergJASAEL,HirschbergAIAA1,HirschbergAIAA2}. For the purposes of this study $Q_E$ is of greater interest as it can be used to generate entropy patches, as will be discussed in \textbf{\S Entropy-patch choked-nozzle interaction simulations}.

The computational procedure used to carry out the numerical study is the same as reported by Elbakly \cite{Elbakly_report}. It consists of three sequential steps:

\begin{enumerate}
    \item Generation of a computational mesh (detailed in \textbf{\S Mesh generation}).
    \item Establishing steady choked-nozzle base flow (expanded upon in \textbf{\S Establishing choked-nozzle base flow}).
    \item Running unsteady entropy-patch-choked-nozzle-interaction (ECNI) simulations (discussed in \textbf{\S Entropy-patch choked-nozzle interaction simulations}).
\end{enumerate}

\subsection{Mesh Generation}\label{sec:grid_generation}

In this section, details regarding mesh generation are provided for the convergent-divergent-nozzle configuration used for the present study. 

\begin{figure}[h]
    \centering
    \includegraphics[width=\linewidth]{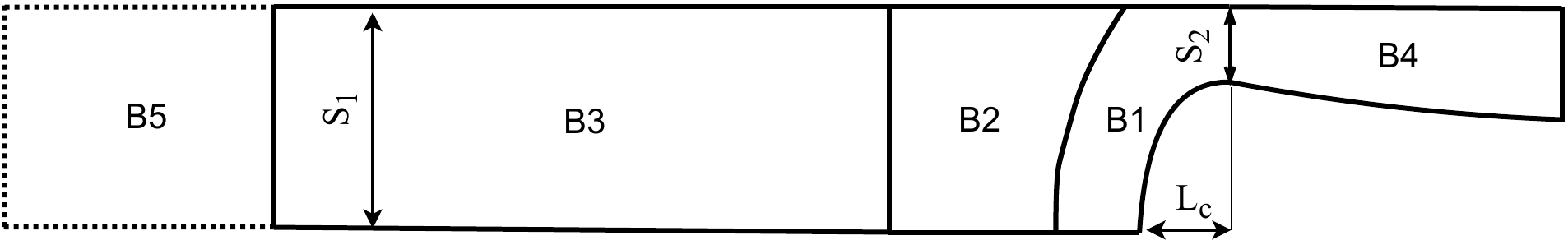}
    \caption{Schematic of multi-block numerical domain with $S_1/S_2 = 3$}
    \label{fig:numerical_domain}
\end{figure}

The schematic of the numerical domain used can be seen in Fig. \ref{fig:numerical_domain}. The domain utilizes a multi-block system where blocks B1, B2, B3, and B4 are used for carrying out the numerical computations. Block B5 is a (passive) 1-D extraction region used to ensure only planar acoustic waves are recorded---note that block B5 is a post-processing block and does not affect any of the adjacent blocks \cite{EIA}. 

The upstream channel is defined by block B3 with a height $S_1=1\; \mathrm{m}$, in which entropy patches are generated and convected downstream. Block B2 acts as a transition zone to minimize cell stretching and skew. The convergent part of the nozzle (nozzle inlet) is Block B1. The nozzle inlet curve was generated using a Henrici transform \cite{Henrici} as previously done by Hirshberg\cite{hirschberg_PHD}, with a contraction ratio $S_1/S_2 = 3$ and length ratio $L_c/S_1 = 1/2$. Lastly, downstream on the contraction block, B4 acts as the diffuser. Note that the geometry of the diffuser is not of high importance as only choked-nozzle flow is considered, implying a supersonic condition in the diffuser preventing the travel of information upstream.  

To establish appropriate levels of grid resolution, a discretization error study was carried out by Elbakly \cite{Elbakly_report}. In the study, base flows were established using a four-stage second-order accurate (4,2) Runge-Kutta scheme for time discretization and a second-order total-variation-diminishing (TVD) Roe approximate Riemann solver with a van Leer limiter for space discretization. ECNI simulations were done using a five-stage second-order accurate (5,2) Runge-Kutta scheme for time discretization and a second-order total-variation-diminishing (TVD) Roe approximate Riemann solver with a van Leer limiter for space discretization. Elbakly~\cite{Elbakly_report} established that using a grid resolution of 30 points per circular entropy spot radius $R_s$ or rectangular block half-width $W_s$ gives a discretization error of 2.2\% at an observed order of accuracy of 1.9 at a distance $42R_s/W_s$ away from the nozzle inlet. Ergo, for the current study meshes were generated using a base value of 30 points per length scale of the entropy patch.

\subsection{Establishing choked-nozzle base flow}\label{sec:establishing_base_flow}

In this section, the computational procedure used to establish a base flow with a choked-nozzle condition is briefly discussed, for an expansive explanation the reader is referred to Elbakly's report \cite{Elbakly_report}.

Elbakly's computation procedure \cite{Elbakly_report} makes use of a pre-established steady base flow to initialize the flow field for the unsteady ECNI simulation. To establish this base flow two separate numerical runs are required. An initial condition run is performed to establish a choked-nozzle flow, followed by an intermediate run to homogenize the flow field and ensure the passing of transients arising from changing boundary conditions between the initial condition run and the ECNI run. 

For the initial condition run the flow field was initialized by setting the blocks upstream of the nozzle throat (B1, B2, B3, \& B5) to have density $\rho = 1~\mathrm{kg}\cdot \mathrm{m^{-3}}$, specific heat ratio $\gamma = 1.4$, pressure $ p = c^2\rho/\gamma \approx 0.7~\mathrm{kg\cdot s^{-2}\cdot m^{-1}}$, and velocity $\mathbf U \equiv (U_\mathrm{des},0)^T \approx (0.2,0)~\mathrm{m\cdot s^{-1}}$. $U_\mathrm{des}$ was the imposed inlet velocity needed to establish a choked-nozzle flow for the given contraction ratio\cite{thompson1972compressible}. Block B4 downstream of the nozzle throat was initialized by setting $\rho = 1.0~\mathrm{kg\cdot m^{-3}}$,  $\gamma = 1.4 $,  $ p = c^2\rho/2\gamma \approx 0.4~\mathrm{kg\cdot s^{-2}\cdot m^{-1}}$, and $\mathbf U \equiv (1,0)^T~\mathrm{m\cdot s^{-1}}$. Note that for the whole domain the local speed of sound $c$ was assumed to be one as per the definition of pressure. Block B4 was set to have a pressure two times lower than that of the upstream blocks to ensure no shockwaves are formed in the diffuser [The formation of shockwaves does not negatively affect the upstream recorded signal as one has a sonic condition at the throat preventing the travel of flow features upstream. However, the formation of shockwaves negatively impacts the computation time by reducing the allowable time step].

For the initial condition run, a constant inflow condition was set on the left bound of blocks B3 and B5 maintaining  $\rho = 1.0~\mathrm{kg\cdot m^{-3}}$, $\mathbf U \equiv (1,0)^T~\mathrm{m\cdot s^{-1}}$ and  $c = 1~\mathrm{m\cdot s^{-1}}$. Pressure relief surfaces on the upper and lower bounds of block B3 were imposed, by setting $R \equiv p^{-}/p^{+} = 0$. Wall boundary conditions were imposed on the lower bounds of blocks B1, B2, and B4. On the upper bounds of blocks B1, B2, and B4 symmetry boundary conditions were imposed. Non-reflective boundary conditions are applied on the right bound on block B4 to imitate anechoic termination. A 1-D boundary condition on the upper and lower bound of block B5 was imposed to classify the block as a post-processing block. Extraction boundary conditions were applied on the interface between block B3 and B5 to ensure a unidirectional transfer of information from block B3 to B5. Connection boundary conditions were applied to the remaining block interfaces to allow for the transfer of data between blocks.

After the domain was fully defined, simulations were initiated and ran until they fully converged. A Roe-TVD scheme with a Van Leer limiter was used for spatial discretization. Temporal discretization was done by means of a second-order accurate four-stage (4,2) Runge-Kutta scheme with alpha coefficients (0.240, 0.375, 0.5, 1.0) and a max Courant limit of one, with artificial dissipation running on the first stage. Time marching was done using a non-time-accurate method as one is interested in the steady state solution.

The converged solution of the initial condition run was used to initialize the domain for the intermediate run. For the intermediate run, some of the boundary conditions are altered. For instance, the constant inflow boundary conditions imposed on the left bounds of blocks B3 and B5 are replaced by non-reflective boundary conditions to imitate anechoic terminations [Note that removing inflow boundary conditions does not halt flow in the numerical domain as EIA makes use of the Euler equations which are lossless]. Furthermore, the pressure relief wall conditions applied to the upper and lower bounds of block B3 are replaced with symmetry \& wall symmetry boundary conditions. The intermediate run was matched to full convergence using the same numerical methods as used for the initial condition run. 

\begin{figure}[h]
    \centering
\begin{subfigure}
    \centering
    \hspace{10pt}
    \includegraphics[width=0.85\textwidth]{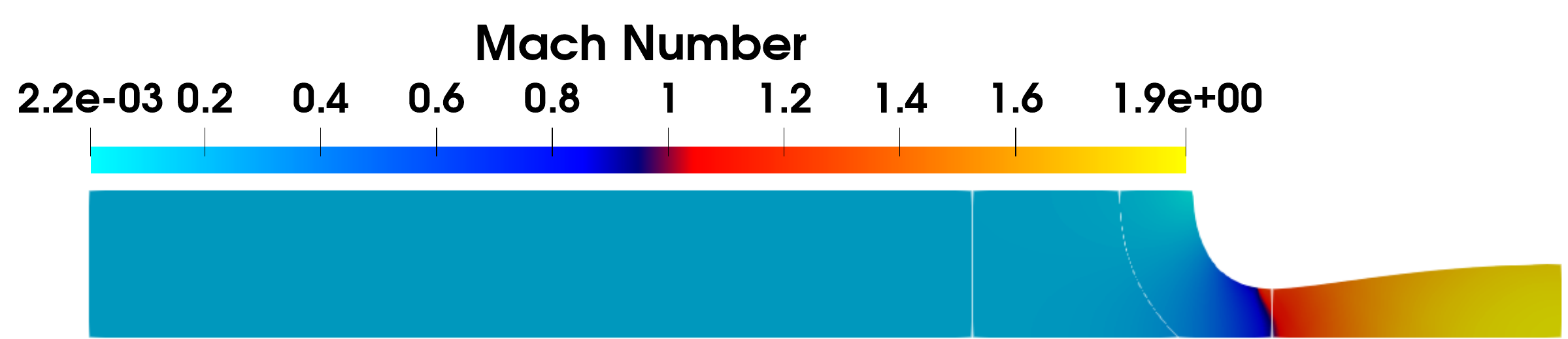}
\end{subfigure}
\begin{subfigure}
    \centering
    \includegraphics[width=0.89\textwidth]{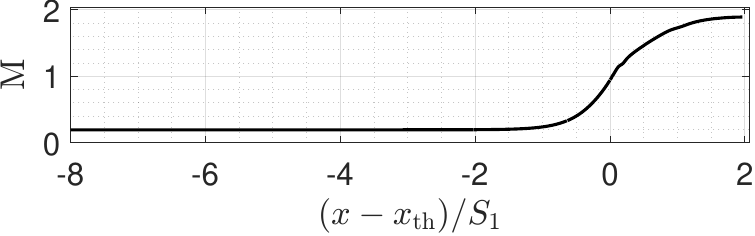}
\end{subfigure}
    \caption[]{Mach field of numerical domain resulting from intermediate run accompanied by the mach profile along the symmetry line ($M_u\approx0.2$).}
    \label{fig:mach_profile}
\end{figure}

The flow field resulting from the intermediate run and the associated Mach profile along the symmetry line is shown in Fig. \ref{fig:mach_profile}. One notes that the upstream chamber is uniform and has a mach number $M_u\approx0.2$. 

\subsection{Entropy-patch choked-nozzle interaction simulations}\label{sec:ECNI_run}
In this section, the computational procedure used to carry out ECNI runs is discussed. Additionally, the most important aspects of entropy-patch generation are expanded upon.

The ECNI run uses the flow field resulting from the intermediate run to initialize the flow field, using the same boundary conditions. The differences between the intermediate run and the ECNI run are as follows: 

\begin{itemize}
    \item The placement and use of a data-recording probe in the 1-D extraction region at a distance $10S_1$ from the nozzle inlet. This pressure probe was set to record density, pressure, velocity, and temperature.
    \item The numerical methods used to resolve the flow field. The ECNI run makes use of a Roe-TVD scheme with a Van Leer limiter for spatial discretization. For temporal discretization, a second-order accurate five-stage (5,2) Runge-Kutta scheme with alpha coefficients (0.125, 0.1666, 0.375, 0.5, 1.0) with artificial dissipation set to run only on the first two stages. A time-accurate method was used with a maximum Courant limit of two.
    \item The generation of entropy patches on top of the background flow. This is expanded upon in the remainder of this subsection.
\end{itemize}

\begin{figure}[ht]
    \centering
    \begin{subfigure}
    \centering
    \vspace{-2pt}
        \includegraphics[width =0.49\textwidth]{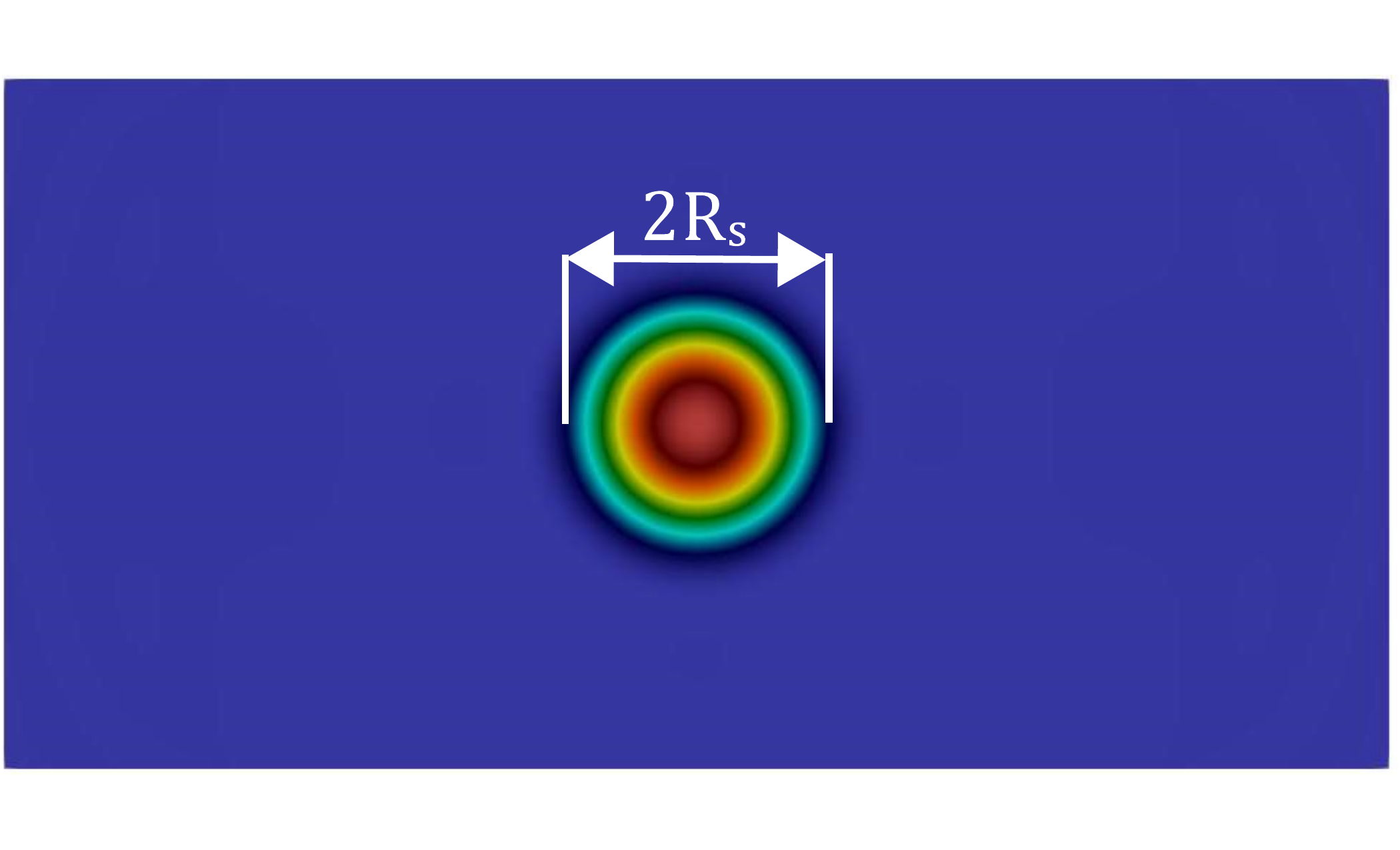}
    \end{subfigure}%%
    \begin{subfigure}
    \centering
        \includegraphics[width =0.49\textwidth]{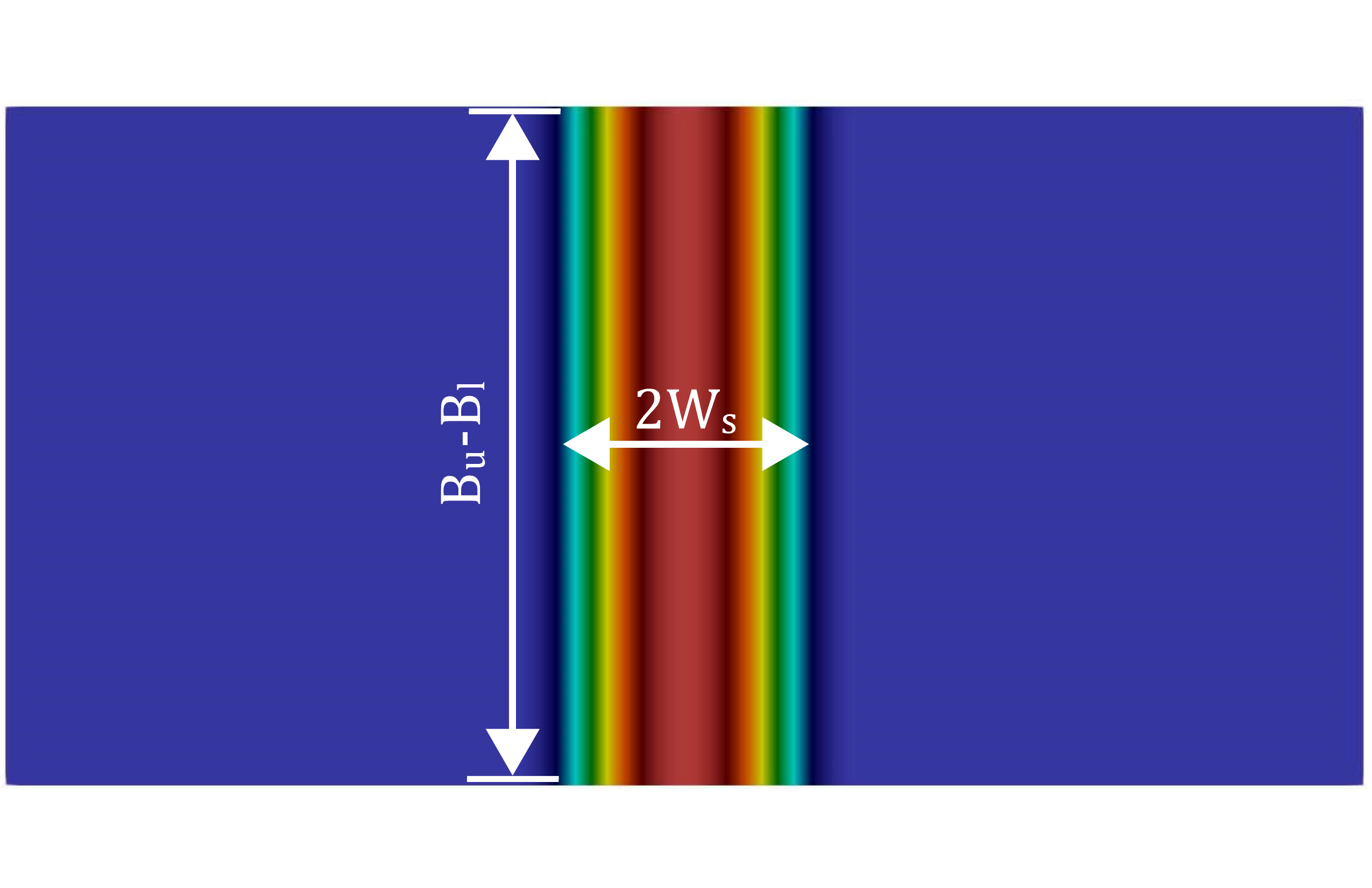}
    \end{subfigure}
    \caption[Entropy field showing fully mature entropy patches]{Entropy field showing fully mature entropy patches; entropy spot (right), entropy block (left).}
    \label{fig:ent_b_s}
\end{figure}

As mentioned above, entropy patches were generated using the $Q_E$ term in Eq. \ref{eq:EnergyEuler}. In EIA, this is done by means of direct energy injection around a moving reference point [the reference point is normally set to move with background flow] with a local distribution of $Q_E$ in the energy balance equation, which influences local entropy generation. One can generate two types of entropy patches: entropy spots (circular patches) and blocks (rectangular slug-like patches). For both these patches we identify a characteristic streamwise length scale, see Fig. \ref{fig:ent_b_s}. In the case of an entropy spot it is the radius of the spot $R_s$, and for an entropy block it is the half-width of the block $W_s$. 

$Q_E$ is defined as a function of the generation amplitude $A_\mathrm{gen}$ and the distance from the center of the generation region $\zeta$ in the direction of the length scale. For entropy spots the source term is defined as follows:

\begin{equation}
    Q_E = \begin{cases}
        \frac{A_\mathrm{gen}}{2} \left( 1 + \cos\left( \pi \frac{\zeta}{R_s} \right) \right) & \text{if } \zeta \in [0, R_s] \\[10pt]
        0 & \forall \; \zeta \notin [0, R_s]
    \end{cases}
\end{equation}

\noindent and for entropy blocks the source term is

\begin{equation}
    Q_E = \begin{cases}
        A_\mathrm{gen} \cos\left(\frac{\pi}{2} \frac{\zeta}{W_s}\right)^2 & \text{if } \zeta \in [0, W_s]|[B_l,B_u] \\[10pt]
        0 & \forall \; \zeta \notin [0, W_s]|[B_l,B_u]
    \end{cases}
\end{equation}

\noindent where $B_l$ and $B_u$ are the lower and upper bounds of the generation region. The generation amplitude is time-dependent and is defined as follows:

\begin{equation}
A_\mathrm{gen} =
\begin{cases} 
\frac{A_\mathrm{max}}{2} \left( 1 - \cos \left( \frac{\pi t}{\tau_{\text{start}}} \right) \right) & \text{if } t \in [0, \tau_{\text{start}}], \\[10pt]
A_\mathrm{max} & \text{if } t \in (\tau_{\text{start}},  \tau_{\text{start}} + \tau_{\max}] \\[10pt]
\frac{A_\mathrm{max}}{2} \left( 1 + \cos \left( \frac{\pi \left( t - (\tau_{\text{start}} + \tau_{\max}) \right)}{\tau_{\text{end}}} \right) \right) & \text{if } t \in \left( \tau_{\text{start}} + \tau_{\max}, \tau_{\text{start}} + \tau_{\max} + \tau_{\text{end}} \right] \\[10pt]
0 & \text{if } t \in \left( \tau_{\text{start}} + \tau_{\max} + \tau_{\text{end}}, t_{\text{end}} \right)
\end{cases}
\end{equation}

\noindent where $t$ is the time, $A_\mathrm{max}$ the maximum generation amplitude, $\tau_\mathrm{start}$ the ramp-up time of entropy generation, $\tau_\mathrm{max}$ the generation time at max amplitude, and $\tau_\mathrm{end}$ the wind-down time. Note that $R_s$, $W_s$, $B_u$, $B_l$, $A_\mathrm{max}$, $\tau_\mathrm{start}$, $\tau_\mathrm{max}$, $\tau_\mathrm{end}$ are user-set variables.

All entropy patches (spots and blocks) used for the current study were generated with the same $A_\mathrm{gen}/A_\mathrm{max}$ profile; viz., $\tau_\mathrm{start} = \tau_\mathrm{end}=6~\mathrm{s}$ and $\tau_\mathrm{max}=3~\mathrm{s}$. Entropy spots were generated with their reference point moving along the symmetry line. The entropy blocks were generated with their upper and lower bounds coinciding with the upper and lower bounds on the numerical domain. Furthermore, it was ensured that the entropy patches were fully mature before they left the generation block B3 such that there was no overlap between the pressure perturbations due to entropy generation and entropy-patch choked-nozzle interaction.

\section{Results \& discussion}

\subsection{Use of reduced-order models for scaling analysis}\label{sec:res_reduced_order_models}

\subsubsection{Matching-condition model-based scaling}

The matching-condition model requires the relative excess density ($\rho^\prime_e/\rho_u$) of the entropy patch. In the cases considered here, entropy patches have an excess density $\rho^\prime_e<0$. The relative excess density was approximated as follows: $(\rho^\prime_{e}/\rho_u)\simeq-|\rho^\prime_{e}/\rho_u|_{\max}$.

For each simulation $|\rho^\prime_{e}/\rho_u|_{\max}$ is determined using an EIA functionality developed by Elbakly~\cite{Elbakly_report}. Its value is then substituted in Eq. \ref{eq:matching_conditon_model} to determine $p_\mathrm{matching}'$, which is then used to scale the simulation results. This was done to determine if the generated upstream acoustic response is due to a quasi-steady mechanism. I.e., it was, on a case-by-case basis, used to establish whether or not the ECNI results were in the matching-condition modeling regime.  

\subsubsection{Inertial model based scaling}

\begin{figure}[t]
    \centering
    \includegraphics[width=0.75\linewidth]{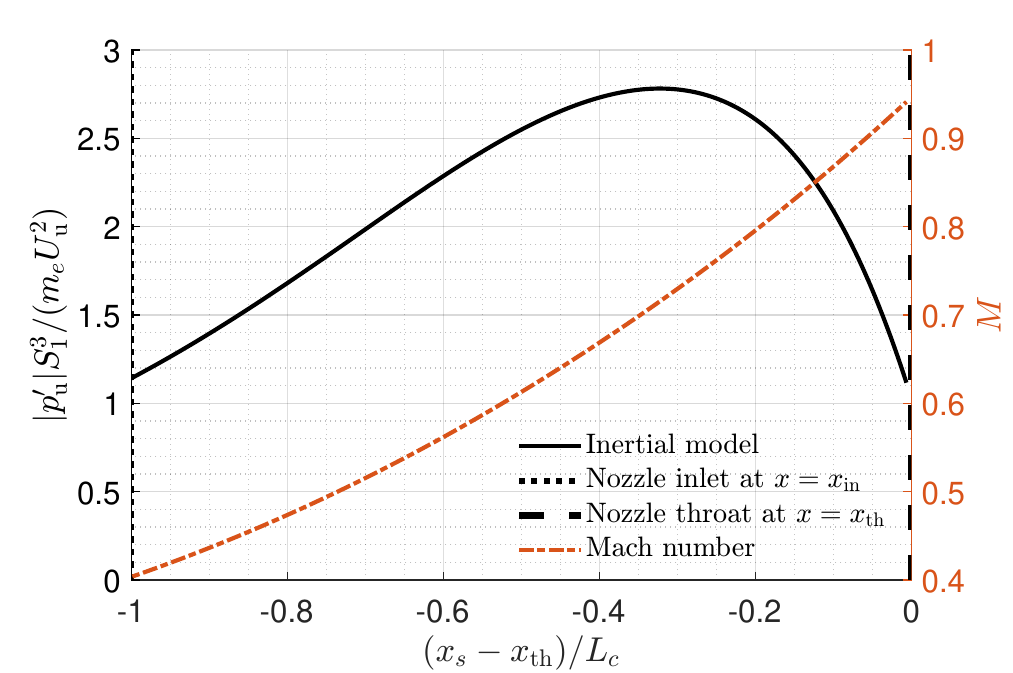}
    \caption{Dimensionless upstream observed acoustic response $|p_u'|S_1^3/(m_eU_u^2)$ obtained from the inertial model (right axis) and mach number $M$ along the symmetry line (left axis) as a function of the dimensionless source position $(x_s-x_\mathrm{th})/L_c$ in the convergent part of the nozzle.}
    \label{fig:inertial_model}
\end{figure}

The convective acceleration $u(\mathrm{d}u/\mathrm{d}x)$, needed as an input for the inertial model, was extracted along the symmetry line from the flow field established during the intermediate run. Using $u(\mathrm{d}u/\mathrm{d}x)$, we determined the dimensionless upstream acoustic response $|p_u'|S_1^3/(m_eU_u^2)$ as a function of dimensionless sound-source location $(x_s-x_\mathrm{th})/L_c$ in the convergent part of the quasi-1-D nozzle (see Fig. \ref{fig:inertial_model}). One notes that there is a maximum in dimensionless acoustic response at $(x_s-x_\mathrm{th})=0.32L_\mathrm{c}$. Going forward, we will refer to this predicted maximum amplitude as $p_\mathrm{inertial}'$. For a fixed nozzle geometry, which we consider here,  $p_\mathrm{inertial}'$ is determined by the excess mass $m_e$ carried by the entropy patch. Thus, one can estimate the $p_\mathrm{inertial}'$ for a given simulation by extracting $m_e$. This is done using a purposely developed EIA functionality \cite{Elbakly_report}. Note that EIA is a planer 2-D code meaning that the value of the excess mass obtained from it $m_{e,EIA}$ is in $kg\cdot m^{-1}$ and should be scaled by the depth of the computational domain (Fig. \ref{fig:computational_domain}), $m_e = 2S_1 m_{e,EIA}$, to be used in the inertial model.

To determine whether the upstream acoustic-response amplitude $|p_u'|$ resulting from an ECNI run is due to the acceleration of the patch, it is scaled by $p_\mathrm{inertial}'$. In other words, $|p_u'|/p_\mathrm{inertial}'$ is computed for each ECNI run to determine whether or not the relevant sound-production mechanism is in the inertial modeling regime.

\subsection{Determination of the modeling regimes: Effect of entropy patch shape and size on the upstream acoustic response scaled by reduced-order models}

To judge whether or not it is essential to model acceleration explicitly for sound production in a given ECNI run, a series of numerical simulations were executed to record the upstream acoustic response due to entropy-patch-coked-nozzle interaction. Entropy patches with a dimensionless streamwise length scales $L_\mathrm{s}/S_2 = \{0.15, 0.3, 0.6, 1.5, 3.0, 4.5, 6.0, 7.5, 9.0, 12.0\}$ were considered, where $L_s$ is the radius $R_\mathrm{s}$ for spots and half-width $W_\mathrm{s}$ for blocks. The results analyzed in this subsection were generated using the same maximum generation amplitude $A_\mathrm{max}=0.03~W\cdot m^{-3}$. We note that the effect of amplitude was investigated as well, however the results of that study are discussed in \textbf{\S Effect of entropy-patch strength on upstream acoustic response prediction by
reduced-order models}. 

\begin{figure}[h]
    \centering
    \includegraphics[width=\linewidth]{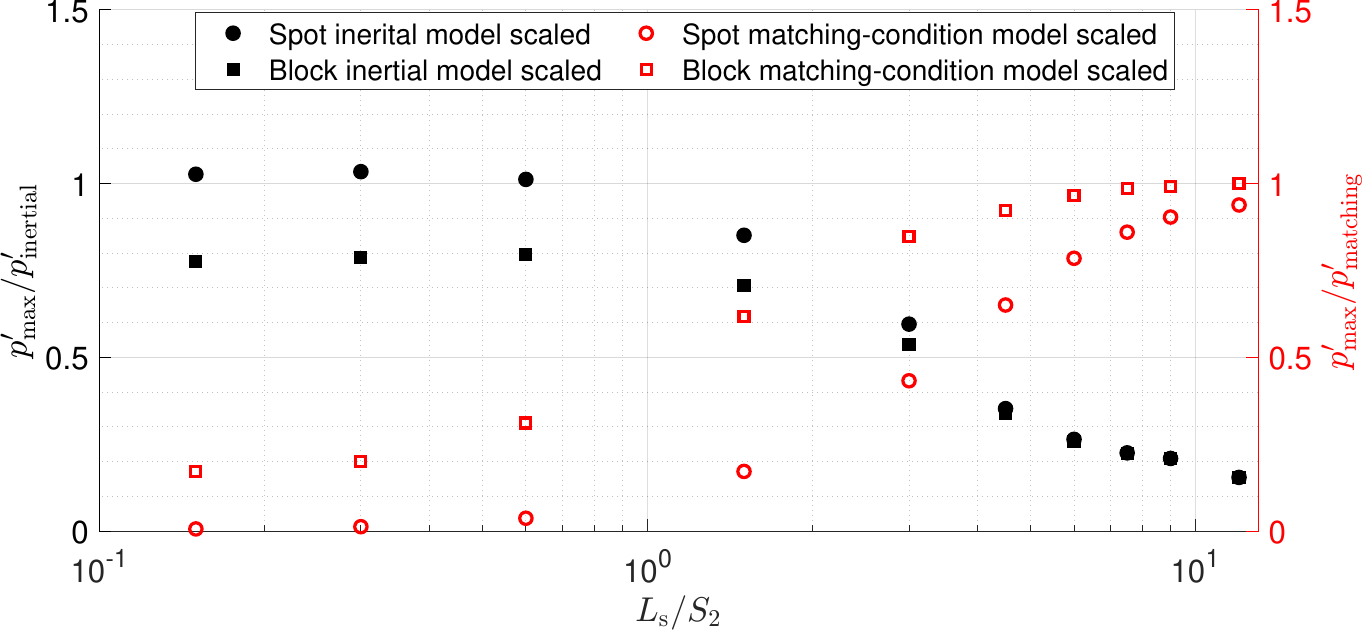}
    \caption{The maximum upstream acoustic response due to entropy-patch chocked nozzle interaction obtained from numerical simulations $p^\prime_\mathrm{max}$ scaled by the inertial model $p^\prime_\mathrm{inertial}$ (left axis) and the matching condition model $p^\prime_\mathrm{matching}$ (right axis) as a function of dimensionless streamwise length scale $L_\mathrm{s}/S_2$, where $L_\mathrm{s}=R_\mathrm{s}$ for entropy spots and $L_\mathrm{s}=W_\mathrm{s}$ for entropy blocks. }
    \label{fig:inertial_vs_matching}
\end{figure}

In Fig. \ref{fig:inertial_vs_matching}, the scaled maximum upstream acoustic response obtained from numerical simulations is plotted as a function of the dimensionless streamwise length scale of the patch $L_s/S_2$, where $S_2$ is half the nozzle-throat height. In black (filled squares/circles): the maximum upstream acoustic response $p_\mathrm{max}'$ scaled by $p^\prime_\mathrm{inertial}$ (left-hand vertical axis). In red (unfilled squares/circles): the maximum upstream acoustic response $p_\mathrm{max}'$ scaled by $p^\prime_\mathrm{matching}$ (right-hand vertical axis). Entropy spots and blocks are represented as circles and squares respectively. 

Two modeling regimes can be identified in Fig. \ref{fig:inertial_vs_matching}; viz., the inertial-modeling regime for $L_\mathrm{s}/S_2\leq1$ and the matching-condition modeling regime for $L_\mathrm{s}/S_2>1$. Indeed, asymptotic behavior is observed in both cases. 

That said, in the case of the matching-condition model entropy blocks compared to entropy spots seem, generally speaking, to be better captured by the model. Conversely, in the case of the inertial model entropy spots seem to a have better representation by the model. We conjuncture that in the case of the matching-condition model this is due to the reduction of the domain to a 1-D line, hence variations in geometry are not captured. For the inertial model it is hypothesized that this difference is due to an entropy patch being considered a \textbf{point particle} during the derivation (Eq. \ref{eq:fx}), for which entropy spots are a better representation. 

These results confirm the preliminary results presented by Kowalski et al.~\cite{Kowalski} with regard to the existence of two distinct modeling regimes; viz., a matching condition modeling regime and an inertial modeling regime. Moreover, the blended-effects regime, where both matching conditions and inertial effects play a role in the production of an upstream acoustic, posited by Kowalski et al.~\cite{Kowalski} remains a viable hypothesis. 

However, here we identified and demonstrated, for the first time, that  $L_\mathrm{s}/S_2$ is an apt dimensionless order parameter. Indeed, $L_\mathrm{s}/S_2$ indubitably allows one to identify by their asymptotes the two matching and inertial regimes. Moreover, this order parameter points to the fact that the size of the entropy patch relative to the throat height is the dominant feature in determining which modeling regime is applicable. E.g., one can confidently assert that for $L_s/S_2\geq 1$, in which case the entropy patch fully occupies the nozzle throat and thus fully changes the thermodynamic state of in nozzle throat: quasi-steady matching condition modeling applies.  Whereas, if the entropy patch does not fully occupy the nozzle throat and the force exerted by the walls of the nozzle due to the acceleration of an entropy patch dominates, one falls under the inertial modeling regime. i.e., $L_s/S_2$ allows one to unambiguously identify these two fundamentally different sound production mechanisms. 

We note that both modeling regimes rely on linearization. I.e., there is a caveat: the above holds provided non-linear effects can be neglected. In the following subsection, we establish the limits of linearization as a viable modeling strategy by means of a relevant dimensionless order parameter dubbed the entropy-patch strength.

\subsection{Determining limits of linearized modeling: Effect of entropy-patch strength on upstream acoustic response prediction by reduced-order models}\label{sec:amplitude_dependance}

To investigate linearization as a viable modeling strategy, a series of numerical simulations were performed using entropy patches with different strengths $e^\prime_e/e_u$, where $e_u$ is the area-specific total energy of the upstream channel defined as 
\begin{equation}
    e_u \equiv \rho_u\left(c_\mathrm{v}T_u+\frac{1}{2}u_u^2\right)
\end{equation}
and $e^\prime_e$ is the area-specific perturbation energy of the entropy patch. We note that entropy patches are generated using energy injection at a (moving) point into the main flow. This allows one to define $e^\prime_e$ as the total energy injected during entropy-patch generation scaled by the area of the entropy patch:

\begin{equation}
    e'_e = \frac{1}{A_e} \iint Q_E(\zeta,t) \, \mathrm{d}A \, \mathrm{d}t.
\end{equation}

With that in mind, two sets of six ECNI runs were executed respectively in the inertial modeling regime with $L_\mathrm{s}/S_2 = 0.3$ and matching condition modeling regime with $L_\mathrm{s}/S_2 = 9.0$. Entropy blocks were used given that the shape of the entropy patch has a negligible effect on the applicable modeling regime. For each regime simulations were carried out with $e^\prime_e/e_u = \{0.07,\allowbreak  ~0.30, ~3.00, ~14.95, ~29.90,~89.71\}$. 

\begin{figure}
    \centering
    \subfigure[Entropy blocks with $L_\mathrm{s}/S_2 = 0.3$]{\includegraphics[width=0.49\textwidth]{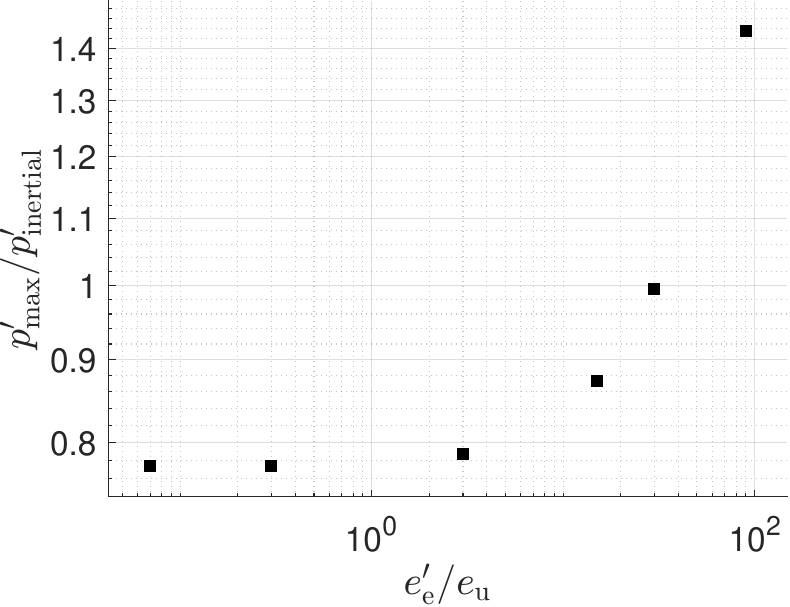}\label{subfig:amp_inertial}}
    \subfigure[Entropy blocks with $L_\mathrm{s}/S_2 = 9.0$]{\includegraphics[width=0.49\textwidth]{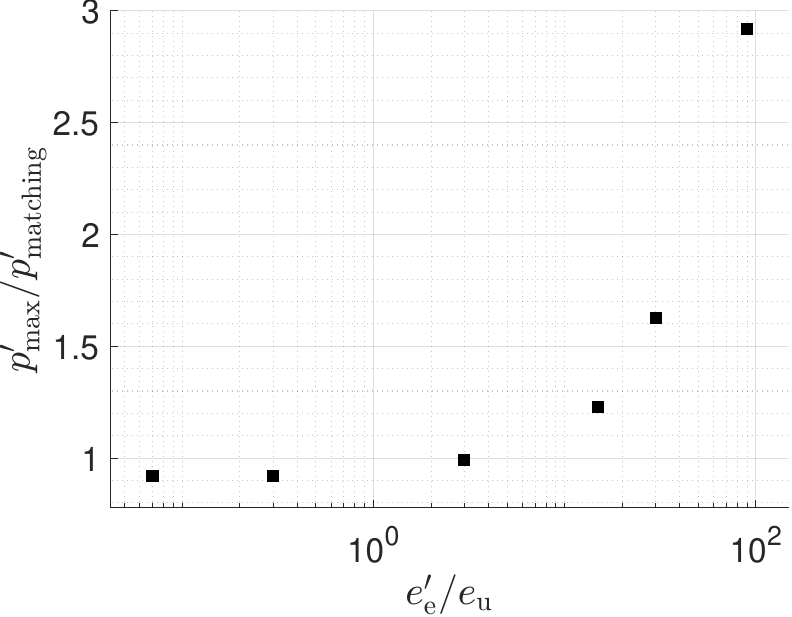}\label{subfig:amp_matching}}
    \caption{The maximum upstream acoustic response due to entropy-patch chocked nozzle interaction obtained from numerical simulations $p^\prime_\mathrm{max}$ scaled by the inertial model $p^\prime_\mathrm{inertial}$ (Fig. \ref{subfig:amp_inertial}) and the matching condition model $p^\prime_\mathrm{matching}$ (Fig. \ref{subfig:amp_matching}) as a function of the ratio of area-specific perturbation energy of the entropy patch to the area-specific total energy of the upstream channel $e^\prime_e/e_u$.}
\label{fig:amp_study}
\end{figure}

The results of these sets of simulations are shown in Fig. \ref{fig:amp_study}. In Fig. \ref{subfig:amp_inertial} $p_\mathrm{max}'$ for $L_s/S_2\leq 1$ (inertial modeling regime) are scaled by $p_\mathrm{inertial}'$ and in Fig. \ref{subfig:amp_matching} $p_\mathrm{max}'$ obtained for $L_\mathrm{s}/S_2>1$ (matching condition modeling regime) by $p^\prime_\mathrm{matching}$. One clearly sees that in both cases for $e^\prime_e/e_u\lesssim1$ a horizontal asymptote is found. One deduces that for $e^\prime_e/e_u\lesssim1$ linearization is an apt modeling strategy.

Furthermore, in both cases the scaled acoustic response approaches a vertical asymptote at $e^\prime_e/e_u\approx10^2$. This vertical asymptote is due to the fact that a physical limit of the system is reached; viz., where one approaches an entropy patch with zero density. One can show this by rewriting $e^\prime_e/e_u$ in terms of the upstream density and the entropy-patch density perturbation as follows (for a detailed derivation refer to \textbf{\S Appendix}):

\begin{equation}\label{eq:e'/e}
\frac{e'_{\mathrm{e}}}{e_{\mathrm{u}}} = \left[ \frac{2}{\gamma (\gamma - 1) M_{\mathrm{u}}^2} + 1 \right] \frac{\rho^\prime_{\mathrm{e}}}{\rho_{\mathrm{u}}}
\end{equation}

\noindent Given that our system has an upstream Mach number of $M_u\approx0.2$ and a specific heat ratio $\gamma=1.4$ one finds $e'_{\mathrm{e}} / e_{\mathrm{u}}\approx90~\rho^\prime_{\mathrm{e}}/\rho_{\mathrm{u}}$. With that in mind, one notes that $e'_{\mathrm{e}} / e_{\mathrm{u}}\approx90$ is possible when the $\rho^\prime_e/\rho_u=\rho_e/\rho_u-1=1$. However, in our ECNI runs $\rho^\prime_e<0$. Thus, $\rho_e$ would have to be zero which is not possible. 

These results show that both the matching condition model and the inertial model are only valid for weak entropy patches $e'_e/e_u\lesssim1$. We note that both models rely on linearization as an essential modeling ingredient. I.e., for $e'_e/e_u\lesssim1$ we are in what we will call a linear-modeling regime.  Here we have ineluctably demonstrated that there are clear limits to the pertinent use of linear modeling. Indeed, for strong patches $e'_e/e_u>1$ neither linear model provides adequate scaling, to wit, nonlinearity is essential.

\section{Conclusion}

A systematic numerical simulation study of entropy-patch choked-nozzle interaction was carried out. The results were analyzed by considering linear reduced-order model-based scaling. This has allowed us to unambiguously confirm the existence of two modeling regimes, to wit, the inertial and (quasi-steady) matching-condition modeling regimes. We have established that the dimensionless order parameter that allows one to determine which modeling regime applies is $L_s/S_2$ (the streamwise length scale divided by half the nozzle throat height). Indeed, our analysis shows that for $L_s/S_2\leq 1$ one finds oneself in the inertial modeling regime, where convective acceleration, which is determined by the nozzle shape, is essential to modeling sound production. For $L_s/S_2 > 1$, the sound production mechanism is in what we have termed the matching-condition modeling regime, where a quasi-steady modeling approach is apt. Sound is produced by changing the thermodynamic state at the choked-nozzle throat. Moreover, specifics of nozzle shape such as radius of curvature, are irrelevant to the matching-condition model as it only relies on the contraction ratio which determines the upstream Mach number.

Two types of entropy patches were considered: circular spots and rectangular (slug-like) blocks. Our results show that the exact shape of the entropy patch has only a marginal effect on the applicable modeling regime needed to study the acoustic response due to entropy-patch choked-nozzle interaction.

We have, in addition, established, by means of a dimensionless order parameter, which we call the entropy-patch strength $e_e'/e_u$, that there are clear limits to the applicability of linearization as a modeling strategy. Indeed, we determined that for $e_e'/e_u>1$ nonlinearity is essential for the description of entropy noise.

\begin{acks}
The authors would like to extend their sincere gratitude to Avraham Hirschberg for his insightful and meaningful comments regarding the present work.
\end{acks}

\bibliographystyle{SageV}
\bibliography{sources.bib}

\begin{thebibliography}{10}
\providecommand{\url}[1]{\texttt{#1}}
\providecommand{\urlprefix}{URL }
\expandafter\ifx\csname urlstyle\endcsname\relax
  \providecommand{\doi}[1]{DOI:\discretionary{}{}{}#1}\else
  \providecommand{\doi}{DOI:\discretionary{}{}{}\begingroup
  \urlstyle{rm}\Url}\fi
\providecommand{\eprint}[2][]{\url{#2}}

\bibitem{strahle_1971}
Strahle WC.
\newblock On combustion generated noise.
\newblock \emph{Journal of Fluid Mechanics} 1971; 49(2): 399--414.
\newblock \doi{10.1017/S0022112071002167}.

\bibitem{Dowling_2}
Dowling AP and Mahmoudi Y.
\newblock Combustion noise.
\newblock \emph{Proceedings of the Combustion Institute} 2015; 35(1): 65--100.
\newblock \doi{10.1016/j.proci.2014.08.016}.

\bibitem{MorgansReview}
Morgans AS and Duran I.
\newblock Entropy noise: A review of theory, progress and challenges.
\newblock \emph{International Journal of Spray and Combustion Dynamics} 2016;
  8(4): 285--298.
\newblock \doi{10.1177/1756827716651791}.

\bibitem{genot2021aluminum}
Genot A.
\newblock Aluminum combustion instabilities: Dimensionless numbers controlling
  the instability in solid rocket motors.
\newblock \emph{Combustion and Flame} 2021; 232: 111563.
\newblock \doi{10.1016/j.combustflame.2021.111563}.

\bibitem{Dotson_5}
Dotson KW, Koshigoe S and Pace KK.
\newblock Vortex shedding in a large solid rocket motor without inhibitors at
  the segmented interfaces.
\newblock \emph{Journal of Propulsion and Power} 1997; 13(2): 197--206.
\newblock \doi{10.2514/2.5170}.

\bibitem{Hulshoff_6}
Hulshoff SJ, Hirschberg A and Hofmans GCJ.
\newblock Sound production of vortex nozzle interactions.
\newblock \emph{Journal of Fluid Mechanics} 2001; 439: 335--352.
\newblock \doi{10.1017/S0022112001004554}.

\bibitem{AnthoineJPP2002}
Anthoine J, Buchlin JM and Hirschberg A.
\newblock Effect of nozzle cavity on resonance in large srm: Theoretical
  modeling.
\newblock \emph{Journal of Propulsion and Power} 2002; 18(2): 304--311.
\newblock \doi{10.2514/2.5935}.

\bibitem{HirschbergJASAEL}
Hirschberg L, Hulshoff SJ, Collinet J et~al.
\newblock Vortex nozzle interaction in solid rocket motors: A scaling law for
  upstream acoustic response.
\newblock \emph{Journal of the Acoustical Society of America} 2018; 144(1):
  EL46--EL51.
\newblock \doi{10.1121/1.5046441}.

\bibitem{HirschbergAIAA1}
Hirschberg L, Hulshoff SJ, Collinet J et~al.
\newblock Influence of nozzle cavity on indirect vortex- and entropy-sound
  production.
\newblock \emph{AIAA Journal} 2019; 57(7): 3100--3103.
\newblock \doi{10.2514/1.J058138}.

\bibitem{HirschbergAIAA2}
Hirschberg L and Hulshoff SJ.
\newblock Lumped-element model for vortex-nozzle interaction in solid rocket
  motors.
\newblock \emph{AIAA Journal} 2020; 58(7): 3241--3244.
\newblock \doi{10.2514/1.J058673}.

\bibitem{Bake_2009}
Bake F, Richter C, Muhlbauer B et~al.
\newblock The entropy-wave generator~(\protect{EWG}): A reference case on
  entropy noise.
\newblock \emph{Journal of Sound and Vibration} 2009; :
  574--598\doi{10.1016/j.jsv.2009.05.018}.

\bibitem{KingsSpray_12}
Kings N and Bake F.
\newblock Indirect combustion noise: noise generation by accelerated vorticity
  in a nozzle flow.
\newblock \emph{International Journal of Spray and Combustion Dynamics} 2010;
  2(3): 253--266.
\newblock \doi{10.1260/1756-8277.2.3.253}.

\bibitem{KingsThesis}
Kings N.
\newblock \emph{Indirect combustion noise: Experimental investigation of the
  vortex sound generation in nozzle flows}.
\newblock PhD Thesis, Technische Universit\"at Berlin, 2015.

\bibitem{HirschbergAIAA4}
Hirschberg L, Bake F, Knobloch K et~al.
\newblock Swirl-nozzle interaction experiments: Influence of
  injection-reservoir pressure and injection time.
\newblock \emph{AIAA Journal} 2021; 59(7): 2806--2810.
\newblock \doi{10.2514/1.J060291}.

\bibitem{Domenico_2021}
De~Domenico F, Rolland E, Rodrigues J et~al.
\newblock Compositional and entropy indirect noise generated in subsonic
  non-isentropic nozzles.
\newblock \emph{Journal of Fluid Mechanics} 2021; 910: A5 1--31.
\newblock \doi{10.1017/jfm.2020.916}.

\bibitem{Noiray_2021}
Wellemann M and Noiray N.
\newblock Experiments on sound reflection and production by choked nozzle flows
  subject to acoustic and entropy waves.
\newblock \emph{Journal of Sound and Vibration} 2021; 492: 115799.
\newblock \doi{10.1016/j.jsv.2020.115799}.

\bibitem{HirschbergEXIF}
Hirschberg L, Bake F, Knobloch K et~al.
\newblock Swirl-nozzle interaction experiment: Quasi-steady model based
  analysis.
\newblock \emph{Experiments in Fluids} 2021; 62(175): 1--16.
\newblock \doi{10.1007/s00348-021-03271-y}.

\bibitem{HirschbergJSV_2}
Hirschberg L, Bake F, Knobloch K et~al.
\newblock Experimental investigations of indirect noise due to modulation of
  axial vorticity and entropy upstream of a choked nozzle.
\newblock \emph{Journal of sound and vibration} 2022; 532.
\newblock \doi{10.1016/j.jsv.2022.116989}.

\bibitem{MC}
Marble FE and Candel SM.
\newblock Acoustic disturbance from gas non-uniformities convected through a
  nozzle.
\newblock \emph{Journal of Sound and Vibration} 1977; 55: 225--243.
\newblock \doi{10.1016/0022-460X(77)90596-X}.

\bibitem{Ffowcs_Williams_howe_1975}
Ffowcs~Williams JE and Howe MS.
\newblock The generation of sound by density inhomogeneities in low mach number
  nozzle flows.
\newblock \emph{Journal of Fluid Mechanics} 1975; 70(3): 605--622.
\newblock \doi{10.1017/S0022112075002224}.

\bibitem{LeykoEtAl}
Leyko M, Moreau S, Nicoud F et~al.
\newblock Numerical and analytical modeling of entropy noise in a supersonic
  nozzle with a shock.
\newblock \emph{Journal of Sound and Vibration} 2011; 330(16): 3944--3958.
\newblock \doi{https://doi.org/10.1016/j.jsv.2011.01.025}.

\bibitem{GentilEtAl_AIAA}
Gentil Y, Daviller G, Moreau S et~al.
\newblock Multispecies flow indirect-noise modeling re-examined with a
  helicopter-engine application.
\newblock \emph{AIAA Journal} 2024; : 1--14\doi{10.2514/1.J063328}.

\bibitem{Kowalski}
Kowalski K, Hulshoff SJ, Str\"{o}er P et~al.
\newblock Entropy-patch-choked-nozzle interaction: Quasi-steady-modeling-regime
  limits probed.
\newblock In \emph{30th AIAA/CEAS Aeroacoustics Conference (2024)}, volume
  2009. American Institute of Aeronautics and Astronautics.
\newblock \doi{10.2514/6.2024-3113}.

\bibitem{KowalskiMScThesis}
Kowalski K.
\newblock \emph{Entropy-patch choked-nozzle interaction: matching-condition and
  inertial-effects regimes mapped \& preliminary investigations of amplitude
  effects}.
\newblock Master's Thesis, University of Twente, 2024.

\bibitem{EIA}
Hulshoff SJ.
\newblock \emph{EIA an Euler Code for Internal Aeroacoustics: method
  description and user's guide}.
\newblock Faculty of Aerospace Engineering, Delft University of Technology,
  Delft, the Netherlands, 2016.

\bibitem{Elbakly_report}
Elbakly K.
\newblock Euler computations of internal aeroacoustics: Development and
  application of numerical solution techniques for indirect combustion noise.
\newblock Technical report, TU Delft, 2024.
\newblock \doi{10.13140/RG.2.2.20370.36807}.

\bibitem{FriedrichBakeUlfMichel}
Bake F, Michel U and Roehle I.
\newblock Investigation of entropy noise in aero-engine combustors.
\newblock \emph{Journal of Engineering for Gas Turbines and Power} 2006;
  129(2): 370--376.
\newblock \doi{10.1115/1.2364193}.

\bibitem{hirschberg_PHD}
Hirschberg L.
\newblock \emph{{Low order modeling of vortex driven self-sustained pressure
  pulsations in solid rocket motors}}.
\newblock Theses, {Universit{\'e} Paris Saclay (COmUE)}, 2019.

\bibitem{thompson1972compressible}
Thompson P.
\newblock \emph{Compressible-fluid Dynamics}.
\newblock Advanced engineering series, McGraw-Hill, 1972.
\newblock ISBN 9780070644052.

\bibitem{Curle}
Curle N.
\newblock The influence of solid boundaries upon aerodynamic sound.
\newblock \emph{Proc Roy Soc A} 1955; 231: 505--514.
\newblock \doi{https://doi.org/10.1098/rspa.1955.0191}.

\bibitem{Pierce}
Pierce AD.
\newblock \emph{Acoustics: an introduction to its physical principles and
  applications}.
\newblock Melville, New York, USA: Acoustical Society of America, 1994.

\bibitem{Henrici}
Henrici P.
\newblock \emph{Applied and Computational Complex Analysis}, volume~I.
\newblock NY, USA: Wiley-Interscience, 1974.

\bibitem{Howe_1998}
Howe MS.
\newblock \emph{Acoustics of Fluid-Structure Interactions}.
\newblock Cambridge Monographs on Mechanics, Cambridge University Press, 1998.
\newblock p.~81.

\end{thebibliography}

\section{Appendix}

\subsection{A relation for the acoustic power emitted from the sound source}\label{append:Derivation_of_acoustic_power}

A general expression for the emitted acoustic power $\Phi$ through a surface of area $A$ is 

\begin{equation}
    \Phi = IA
\end{equation}

\noindent where $I$ is the acoustic intensity. Howe \cite{Howe_1998} showed that the acoustic intensity of an irrotational homentropic (quasi-steady-1D) flow is 

\begin{equation}
    I = (\Bar{\rho}u^\prime+\rho^\prime\Bar{u})B^\prime
\end{equation}
\noindent where $\rho$ is the density, $u$ is the velocity, and $B$ is the total enthalpy while the bar indicates mean-flow quantities and the prime indicates perturbations. $B^\prime$ can be expressed in terms of pressure perturbation $p^\prime$, $\Bar{\rho}$, $\Bar{u}$, and $u^\prime$, as follows:

\begin{equation}\label{eq:b_minus}
    B^\prime = \frac{p^\prime}{\Bar{\rho}}+\Bar{u}u^\prime.
\end{equation}

\noindent Using the relation $u^{\pm}=\pm(p^\pm/\Bar{\rho}\Bar{c})$ and only considering upstream-traveling waves, the total enthalpy fluctuations can be written as 

\begin{equation}\label{eq:b_minus}
    B^- = \frac{p^-}{\Bar{\rho}}-\frac{p^-\Bar{u}}{\Bar{\rho}\Bar{c}} = \frac{p^-}{\Bar{\rho}}\left(1-M\right)
\end{equation}
\noindent where $M$ is the unperturbed flow Mach number.  Using Eq. \ref{eq:b_minus}, and rewriting the velocity and density perturbations, the acoustic intensity of upstream traveling acoustic wave can be expressed as 
\begin{equation}
    I^- = \left(\frac{-p^-}{\Bar{c}}+\frac{p^-\Bar{u}}{\Bar{c}^2} \right)\frac{p^-}{\Bar{\rho}}\left(1-M\right) = -\frac{(p^-)^2}{\Bar{\rho}\Bar{c}}\left(1-M\right)^2.
\end{equation}

\noindent this allows one to express the acoustic power of an upstream traveling acoustic wave as follows
\begin{equation}
    \Phi^- = -\frac{A(p^-)^2}{\Bar{\rho}\Bar{c}}\left(1-M\right)^2.
\end{equation}

\subsection{Derivation of Eq. \ref{eq:e'/e}}\label{append:derivation_of_e'/e}
We define the strength of an entropy patch as follows: 

\begin{equation}\label{eq:decomposed_purtubation_energy}
    \frac{e^\prime_\mathrm{e}}{e_u} \equiv \frac{e_\mathrm{e}-e_u}{e_u}.
\end{equation}

If we consider the general definition of area-specific energy 

\begin{equation}
       e = \rho \left(c_\mathrm{v}T+\frac{1}{2}u^2\right)
\end{equation}
and use the ideal-gas law to rewrite $\rho T$ term, one finds:

\begin{equation}
    e = p \left(\frac{c_v}{R}\right)+\frac{1}{2}\rho u^2.
\end{equation}

\noindent Using this relation to rewrite the energy terms in the numerator of  Eq. \ref{eq:decomposed_purtubation_energy}, yields 

\begin{equation}
   \frac{e^\prime_\mathrm{patch}}{e_u} = \frac{\frac{c_v}{R}(p_\mathrm{e}-p_u)+(\frac{1}{2}\rho_\mathrm{e}u_\mathrm{e}^2-\frac{1}{2}\rho_uu_u^2)}{\rho_u(c_vT_u+\frac{1}{2}u_u^2)} 
\end{equation}

\noindent Noting that $p_\mathrm{e}=p_u$ for mature entropy patches and that the entropy patch is convected by the flow ($u_\mathrm{e}=u_u$), this expression can be rewritten to find:

\begin{equation}
    \frac{e^\prime_\mathrm{e}}{e_u} = \frac{\rho_\mathrm{e}-\rho_u}{\rho_u}\frac{\frac{1}{2}u_u^2}{c_vT_u+\frac{1}{2}u_u^2} = \frac{\rho^\prime_\mathrm{e}}{\rho_u}\left[\frac{u_u^2}{2c_vT_u}+1\right].
\end{equation}

\noindent Now, using the definition of the match number $M=u/c$, speed of sound of ideal gases $c = \sqrt{\gamma RT}$, specific heat ratio $\gamma = c_p/c_v$, and $R = c_p-c_v$, after some algebra, we find 

\begin{equation*}
\frac{e'_{\mathrm{e}}}{e_{\mathrm{u}}} = \frac{\rho^\prime_{\mathrm{e}}}{\rho_{\mathrm{u}}}\left[ \frac{2}{\gamma (\gamma - 1) M_{\mathrm{u}}^2} + 1 \right]
\end{equation*}

\noindent which is Eq. \ref{eq:e'/e}.

\end{document}